\documentclass[referee]{raa}
\usepackage{graphicx,times}
\usepackage{natbib}
\usepackage{amssymb,amsmath}
\bibpunct{(}{)}{;}{a}{}{,}

\usepackage[a4paper=true,dvipdfm=true,pagebackref=true]{hyperref}
\hypersetup{pdftitle = The title of my PDF, pdfauthor = My name, pdfsubject= The subject, pdfkeywords = keyword1 keyword2 keyword3}
\hypersetup{colorlinks = true, linkcolor = green, anchorcolor = red, citecolor = blue, filecolor = red, pagecolor = red, urlcolor = red}

\begin{document}

   \title{Triggering star formation by both radiative and mechanical active galactic nucleus feedback
$^*$
\footnotetext{\small $*$ Supported by the National Natural Science Foundation of China.}
}

 \volnopage{ {\bf 2013} Vol.\ {\bf X} No. {\bf XX}, 000--000}
   \setcounter{page}{1}

   \author{Chao Liu\inst{1,2}, Zhao-ming Gan\inst{1}, Fu-guo Xie
      \inst{1}
   }

   \institute{ Key Laboratory for Research in Galaxies and Cosmology, Shanghai Astronomical Observatory,
Chinese Academy of Sciences, 80 Nandan Road, Shanghai 200030, China; {\it cliu@shao.ac.cn} \\
        \and
             University of Chinese Academy of Sciences, 19A Yuquan Road,
             Beijing 100049, China\\
\vs \no
   {\small Received 2012 xx; accepted xx xx}
}

\abstract{
We perform two dimensional hydrodynamic numerical simulations to study the positive active galactic nucleus feedback which triggers, rather than suppresses, star formation. Recently, it was shown by Nayakshin et al. and Ishibashi et al. that star formation occurs when the cold interstellar medium is squeezed by the impact of mass outflow or radiation pressure, respectively. Mass outflow is ubiquitous in this astrophysical context, and radiation pressure is also important if the AGN is luminous. For the first time in this subject, we incorporate both mass outflow feedback and radiative feedback into our model. Consequently, the ISM is shocked into shells by the AGN feedback, and these shells soon fragment into clumps and filaments because of Rayleigh-Taylor and thermal instabilities. We have two major findings: (1) the star formation rate can indeed be very large in the clumps and filaments. However, the resultant star formation rate density is too large compared with previous works, which is mainly because we ignore the fact that most of the stars that are formed would be disrupted when they move away from the galactic center. (2) Although radiation pressure feedback has a limited effect, when mass outflow feedback is also included, they reinforce each other. Specifically, in the gas-poor case, mass outflow is always the dominant contributor to feedback.
}

   \authorrunning{C. Liu et al. }            
   \titlerunning{Triggering of star formation by AGN feedback}  
   \maketitle

%
\section{Introduction}           
\label{sect:intro}
The observed correlations between the mass of the central supermassive black hole (SMBH) and the characteristic properties of the host galaxy (\citealt{Gebhardt+etal+2000}; \citealt{Ferrarese+etal+2000}; \citealt{Marconi+etal+2003}; \citealt{Fabian+2012}) indicate that active galactic nucleus (AGN) feedback is likely to play an important role in galaxy formation and evolution (\citealt{Silk+Rees+1998}; \citealt{Proga+etal+2000}; \citealt{DiMatteo+etal+2005}; \citealt{Ciotti+etal+2009c}; \citealt{Kurosawa+etal+2009b}; \citealt{Fanidakis+etal+2011}). An AGN can influence its environment and/or its host galaxy in various forms, i.e., radiation pressure, radiative heating, jets and winds/outflows. Jets are believed to be responsible for the formation of the X-ray cavities observed in clusters of galaxies (e.g., \citealt{Fabian+etal+2006}). Since these jets are highly collimated, they are prone to drill through a single galaxy. It is reasonable to ignore the jet feedback in this work, because we focus on the the region within several kiloparsecs of the galactic center (refer to \citealt{Gaibler+etal+2012} for jet-triggering star formation).

In the traditional view, it is believed that AGN feedback impacts its host galaxy in a negative way. Namely, the interstellar medium (ISM) around the SMBH is heated up by photoexcitation/photoionization and Compton heating; or it is blown away by radiation pressure and ram pressure of mass outflows. It is believed that these processes inhibit star formation and gas fueling onto the SMBH (e.g., \citealt{Springel+etal+2005}; \citealt{Farrah+etal+2012}; \citealt{Cano-Diaz+etal+2012}; \citealt{Page+etal+2012}). However, the feedback might be positive in terms of triggering star formation in the host galaxy (\citealt{Fabian+2012}). \cite{Santini+etal+2012} reported evidence of a higher average star formation rate in AGN hosts compared to a control sample of inactive galaxies. The level of star formation enhancement is modest ($\sim0.26$ dex at $\sim3\sigma$ confidence level) at low X-ray luminosity ($L_X\lesssim10^{43.5}~\rm{erg~s^{-1}}$) but more pronounced (0.56 dex at $>$ $10\sigma$ confidence level) in the hosts of luminous AGNs. In another interesting work, \cite{Silk+Nusser+2010} proposed that the star formation triggered by the outflow can help boost the momentum rate from $L_{acc}/c$ released by AGN radiation to $\sim2-30~L_{acc}/c$ deposited into the galactic winds (e.g., \citealt{Moe+etal+2009}; \citealt{Sturm+etal+2011}), where $L_{acc}$ is the AGN accretion luminosity and $c$ is the speed of light.

More recently, two theoretical works focused on the physical processes of positive AGN feedback. One was done by \cite{Nayakshin+etal+2012}, who performed a simulation of the quasar feedback on a gas shell that incorporated smoothed-particle hydrodynamics, and found that when the ambient shocked gas cools rapidly, the shocked gas is compressed into thin cold dense shells, filaments and clumps. Some of these high density features are found to be resilient to the feedback, so they are a hotbed for starbursts. In their work, they only considered mass outflow feedback by assuming wind velocity $v_w=0.1~c$ and wind momentum rate $\dot p_w=L_{Edd}/c$, where $L_{Edd}$ is the Eddington luminosity. The other study was an analytic work by \cite{Ishibashi+etal+2012}, who found that the squeezing and compression of the inhomogeneous interstellar medium can trigger star formation within the dusty gas shell that is driven by radiation pressure when the shell expands outward. They explored the shell's escape/trapping condition in the galactic halo for different underlying dark matter potentials. In their picture, new stars form at increasingly larger radii and successively populate the outer regions of the host galaxy. This inside-out growth pattern seems to match the observational fact that the increase in stellar mass has mainly occurred at outer radii since redshift z$\approx$2 (\citealt{vanDokkum+etal+2010}) just after quasar activity peaked.

The main purpose of this paper is to study the possibility of star formation triggered by AGN feedback through grid-based hydrodynamic (HD) simulations. Our main improvement is to include both mass outflow and radiative feedback in our models. We note that although \cite{Nayakshin+etal+2012} studied quasar feedback, radiative feedback, such as radiative heating and radiation pressure, was ignored. However, quasars are so luminous that radiative feedback is the dominant feedback mechanism, and this is usually referred to as quasar mode feedback (e.g., \citealt{Sijacki+etal+2007}) in cosmological simulations. Our radiative feedback is different from that of \cite{Ishibashi+etal+2012} who only considered the radiation pressure on dust. We ignore dust opacity in our models. Instead, we incorporate into account line-force and Thomson scattering to calculate the radiative feedback force. In addition, we take a more complete treatment of radiative heating/cooling, but there is only bremsstrahlung in \cite{Nayakshin+etal+2012} and no radiative heating/cooling in \cite{Ishibashi+etal+2012}. We aim to investigate the positive AGN feedback with our improved models.

In the following section, we describe our modeling of the AGN feedback mechanisms, numerical methods and model assumptions. We analyze the results of our HD numerical simulations in Section \ref{sect:results}. Finally, we provide conclusions and discussion in Section \ref{sect:concl}.

\section{Methods}
\label{sect:Method}
We focus on the inner part of an isolated galaxy in order to reach a relatively higher resolution. The central engine is treated as a point source composed of two components, an accretion disk that radiates at the ultraviolet (UV) band, where its flux is proportional to $cos(\theta)$, and a corona isotropically radiating at the X-ray band. The radiative heating/cooling mechanisms include Compton heating/cooling, X-ray photoionization heating/recombination cooling, bremsstrahlung and line cooling (\citealt{Proga+etal+2000}). We model the galaxy as a singular isothermal sphere with the total density profile $\rho=\sigma^2/(2\pi Gr^2)$, where $\sigma$ is the velocity dispersion and $G$ is the gravitational constant. Correspondingly, the acceleration of gravity is
\begin{equation}
{\bm g}^{\ast,DM}=-\frac{2\sigma^2}{r^2} \hat{\bm r}.
\end{equation}
"Here $\hat{\bm r}$ is a unit vector along the radial direction. The gas density is assumed to be a fraction $f_g$ of the total density, i.e., $\rho_g=f_g\sigma^2/(2\pi Gr^2)$. Then, the gas mass enclosed in $r$ is $M_g=4\pi \int\rho_g r^2dr=2f_g\sigma^2r/G$.

\subsection{AGN feedback Model}\label{subsec:fedbk model}
Both radiative feedback and mechanical feedback are considered. In terms of the radiative feedback, we follow the treatments of \cite{Kurosawa+etal+2009b}. The radiative heating/cooling and radiation pressure gradient force are added to the energy and momentum equations as source terms, respectively. The radiation pressure is taken into account through Thomson scattering and line-force, where the line-force is parameterized by a line-force multiplier (\citealt{Stevens+Kallman+1990}, see also \citealt{Proga+etal+2000}). The acceleration as a result of radiation pressure can be written as
\begin{eqnarray}\label{eq:rad_pressure}
{\bm g}^{rad}~(r,\theta) = ~\frac{\kappa_{es}}{c}\frac{ L_{acc}}{4\pi r^2}
\left[f_\ast\exp{(-\tau_X)}+(1+\mathcal{M})
\times2\left|\cos\theta\right|f_{d}\exp{(-\tau_{UV})}\right] \hat{\bm r},
\end{eqnarray}
where $\kappa_{es}=0.4~{\rm cm^2~g^{-1}}$ is the mass-scattering coefficient for free electrons, $\mathcal{M}$ is the line-force multiplier, $L_{acc}$ is the accretion luminosity ($L_{acc}=\epsilon_{EM}\dot M_{acc}c^2$, where $\dot M_{acc}$ is the BH mass accretion rate and $\epsilon_{EM}$ is the radiation efficiency), $f_\ast$ and $f_d$ are respectively the X-ray and UV flux fraction, and $\tau_X$ and $\tau_{UV}$ are respectively the X-ray and UV optical depth. We assume that the ISM is optically thin to its own radiation and the UV radiation from the central engine. Therefore, only the radial component is in the equation (\ref{eq:rad_pressure}) and $\tau_{UV}=0$.

Considering a gas shell under the irradiation of the central engine with luminosity $L_{acc}$, one can get the critical luminosity by setting the outward radiation pressure gradient equals to the inward gravitational force, which leads to
\begin{equation}\label{eq:Lcrit}
L_c = \frac {4f_gc\sigma^4}{G}=4.6\times 10^{46} \left(\frac{f_g}{0.16}\right) \left(\frac{\sigma}{2\times10^7{\rm cm/s}}\right)^4 {\rm ergs~s^{-1}} .
\end{equation}
When $L_{acc}>L_c$, the gas shell will be blown away. For simplicity, we set $L_{acc}=L_c$ throughout the current work. If we set the mass of the SMBH to $M_{BH}=10^8 M_\odot$, $\sigma=200~{\rm km~s^{-1}}$ and $f_g=0.16$ (gas-rich case), we immediately have $L_{acc}\simeq3.5L_{Edd}$. But for $f_g=10^{-3}$ (gas-poor case), the accretion luminosity is as low as $L_{acc}\simeq2.2\% L_{Edd}$.

For the mechanical feedback, we consider fast mass outflow with outward radial velocity fixed to $v_w=10000$ km/s. There is much compelling observational evidence for the existence of fast outflows (e.g., \citealt{Crenshaw+1997}; \citealt{Kaastra+etal+2000}; \citealt{Hamann+etal+1997}; \citealt{Chartas+etal+2003}; \citealt{Crenshaw+etal+2003}; \citealt{Hamann+etal+2008}; \citealp{Tombesi+etal+2010, Tombesi+etal+2011a, Tombesi+etal+2012}). We add outflow mass, momentum and kinetic energy as source terms to the basic HD equations at the innermost layer (similar to the treatments by \citealt{Ostriker+etal+2010}). The mass outflow rate $\dot M_w$, outflow momentum rate $\dot P_w$ and kinetic energy rate $\dot E_w$ are calculated as follows:
\begin{equation}\label{eq:Mw}
\dot M_w= \eta_w \frac{L_{acc}}{c} \frac{1}{v_w},
\end{equation}
\begin{equation}\label{eq:Pw}
\dot P_w = \dot M_w v_w = \eta_w \frac{L_{acc}}{c},
\end{equation}
\begin{equation}\label{eq:Ew}
\dot E_w = \frac{1}{2}\dot M_w v^2_w = \epsilon_w\dot M_{acc}c^2,
\end{equation}
where $\epsilon_w$ is wind/outflow efficiency and $\eta_w$ is the ratio of mass outflow momentum rate to radiation momentum rate. The definition of $\eta_w$ can be derived from equations (\ref{eq:Pw}) and (\ref{eq:Ew}),
\begin{equation}\label{eq:etaw}
\eta_w \equiv 2\frac{\epsilon_w}{\epsilon_{EM}}\frac{c}{v_w}.
\end{equation}
The radiation efficiency $\epsilon_{EM}$ is set to be 0.1 throughout this paper. The wind/outflow efficiency is not currently well constrained, but the best estimation is in the range $1\times10^{-3}>\epsilon_w>3\times10^{-4}$ (\citealt{Proga+etal+2000}; \citealt{Proga+Kallman+2004}; \citealt{Krongold+etal+2007}; \citealt{Kurosawa+etal+2009c}; \citealt{Ostriker+etal+2010}). If we set $\epsilon_w=5\times10^{-3}$, then $\eta_w=3$. The outflowing mass flux is also assumed to depend on the angle from the polar axis as $\propto \frac{3}{2} sin\theta cos^2\theta$, so that the half-opening angle enclosing half of the input material is $\approx 45^\circ$. In terms of solid angle, this means that the wind is visible from about $1/4$ of the available viewing angles.

\subsection{Numerical Setup}\label{subsec:setup}
We perform two-dimensional (2D) HD numerical simulations with the modified code ZEUS-MP (\citealt{Stone+etal+1992a}; \citealt{Hayes+etal+2006}) in spherical polar coordinates $(r,\theta,\phi)$. The equations including feedback source terms are:
\begin{equation}\label{eq:mass}
   \frac{D\rho}{Dt} + \rho \nabla \cdot {\bf v} = 0,
\end{equation}
\begin{equation}\label{eq:momen}
   \rho \frac{D{\bf v}}{Dt} = - \nabla P + \rho {\bf g}
 + \rho {\bf g}^{\rm rad} ,
\end{equation}
\begin{equation}\label{eq:ener}
   \rho \frac{D}{Dt}\left(\frac{e}{ \rho}\right) =
   -p \nabla \cdot {\bf v} + \rho {\cal{L}} ,
\end{equation}
where $\rho$ is the mass density, $P$ is the gas pressure, ${\bf v}$ is the velocity, $e$ is the internal energy density, $\rho\cal{L}$ is the net heating rate (\citealt{Proga+etal+2000}), ${\bf g}$ is the total gravitational acceleration including the potentials of the SMBH, dark matter and stars (i.e., $g=GM_{BH}/r+g^{\ast,DM}$), and ${\bf g}^{\rm rad}$ is the acceleration due to radiation pressure (see equation (\ref{eq:rad_pressure})). The outflow feedback terms are not present in the above equations; instead, the mass, momentum and energy of the outflow are directly added to the innermost layer of our simulation domain. We adopt an adiabatic equation of state $P=(\gamma-1)e$, and only consider models with the adiabatic index $\gamma=5/3$.

Our simulation domain covers 5pc to 5kpc in the radial direction and from 0 to $\pi$ in the angular direction. There are 192 radial grids in which $(\Delta r)_{i+1}/(\Delta r)_{i}=10^{1/(N_r-1)}$, and $N_r$ denotes the number of the grid points per decade in radius. Here we set $N_r=64$. In order to better resolve the flow near the equator, we adopt angular zones with $(\Delta \theta)_{j+1}/(\Delta \theta)_{j}=0.985$ for $0\leqslant \theta \leqslant \pi/2$, and $(\Delta \theta)_{j}/(\Delta \theta)_{j+1}=0.985$ for $\pi/2\leqslant \theta \leqslant \pi$. The outflow boundary condition is adopted at the outer radial boundary. We use the same boundary condition as \cite{Novak+etal+2011} at the inner radial boundary, i.e., assuming reflecting boundary conditions if the innermost radial velocity is positive. If the innermost radial velocity is negative, we use an outflow boundary condition where all fluid variables are constant across the boundary. In the angular direction, a symmetrical boundary condition is applied at the polar axes.

\begin{figure}
\centering
\includegraphics[width=14cm, angle=0]{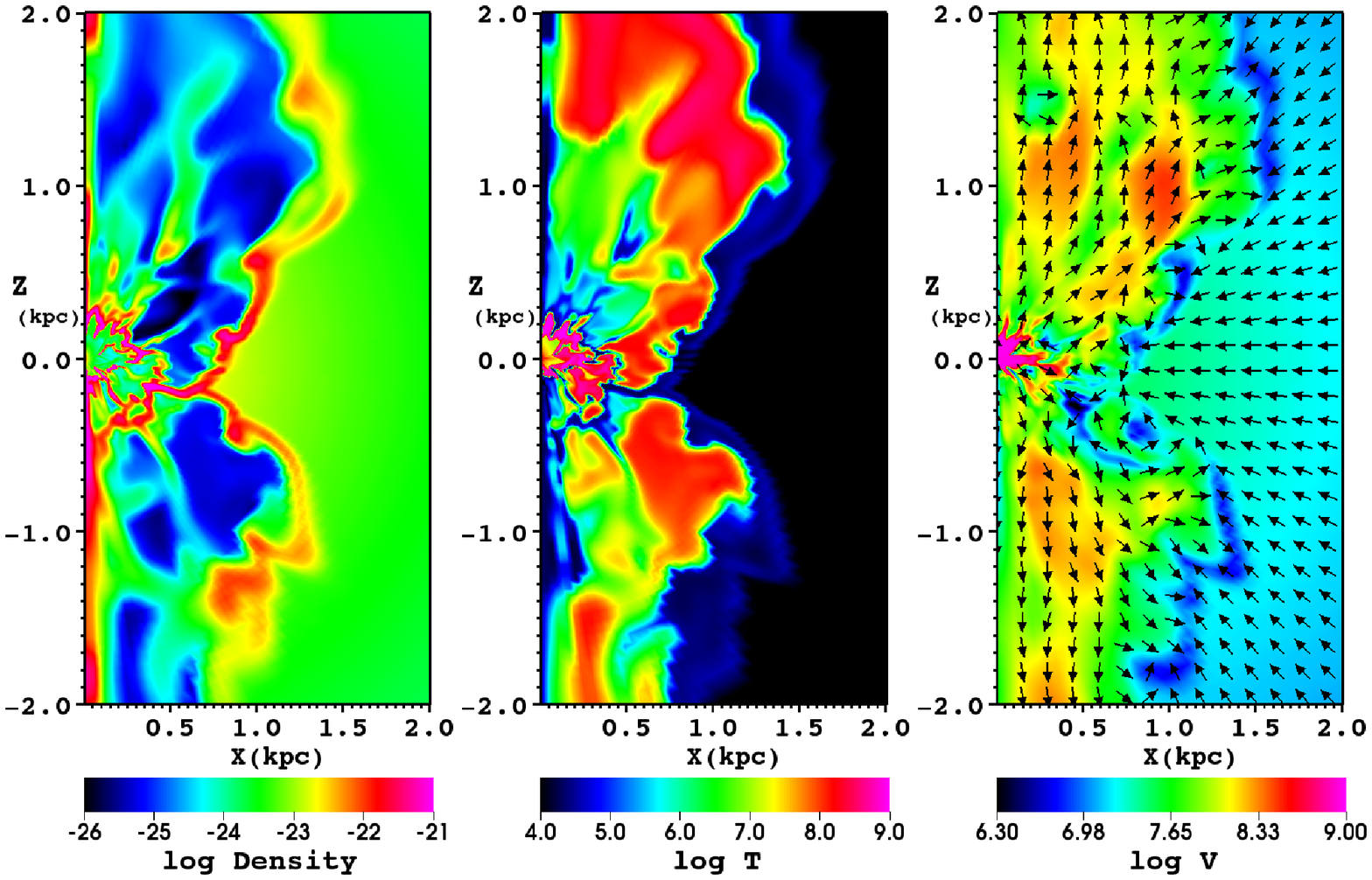}
\includegraphics[width=14cm, angle=0]{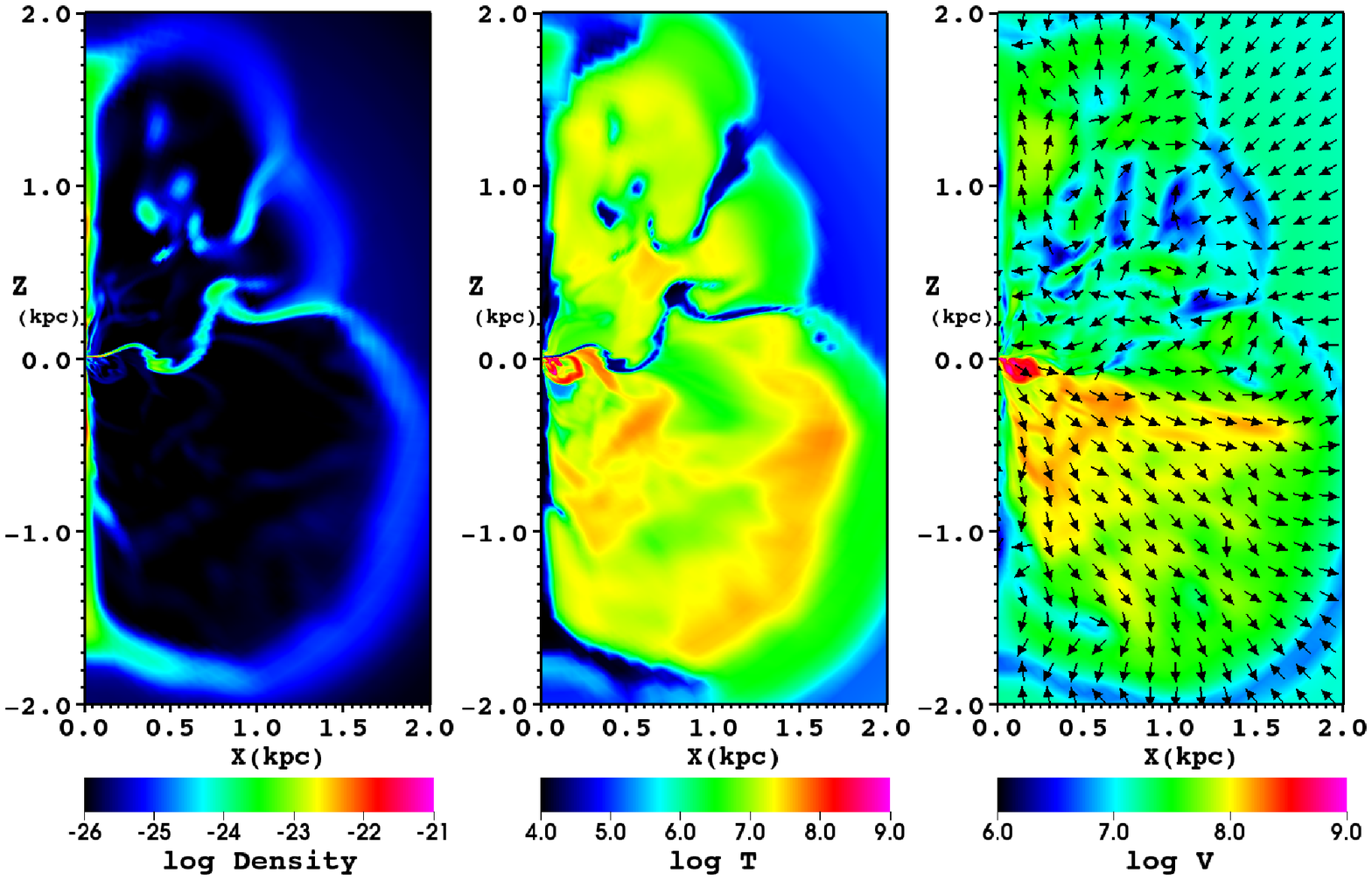}
\caption{{\small Top panels: a gas-rich AMR model. bottom panels: a gas-poor CMR model. From the left to right columns, the quantities are logarithmic gas density (${\rm g~cm^{-3}}$), temperature (K) and velocity magnitude ($\rm cm~s^{-1}$), respectively. The arrows in the right columns show the direction of the velocities. All panels are snapshots taken at $t=4.47$ Myr. The cold dense clumps and filaments in the AMR model are regions with high SFR. These clumps and filaments in which stars form are generated due to Rayleigh-Taylor and thermal instabilities. Most of the stars that form in the clumps and filaments are disrupted when they move away from the center. Clumps and filaments also appear in gas-poor CMR model, which is a new result. See the text for details.}}
\label{fig:Fig1}
\end{figure}

\begin{table}
\bc
\begin{minipage}[]{100mm}
\caption[]{Model summary\label{tab:model_sum}}\end{minipage}
\setlength{\tabcolsep}{2.5pt}
\small
\begin{tabular}{ccccccccccccc}
  \hline
  \hline
{\footnotesize Model} &  {\footnotesize Model} & $f_g{\rm ^a}$   & $\epsilon_w{\rm ^b}$ &  {\footnotesize Radiation Pressure} & {\footnotesize Outflow Feedback} & $t_{kpc}{\rm ^c}$ \\
{\footnotesize Number}&        &         &                 &      {\footnotesize Feedback}       &                  &  {\footnotesize Myr}       \\
  \hline\noalign{\smallskip}
1     &  {\footnotesize AMR}  & 0.16     & $5\times10^{-3}$&  $\surd$       & $\surd$       & 1.92        \\
2     &  {\footnotesize AR}   & 0.16     & $5\times10^{-3}$&  $\surd$       & $\times$      & No          \\
3     &  {\footnotesize AM}   & 0.16     & $5\times10^{-3}$&  $\times$      & $\surd$       & 2.37        \\
  \hline\noalign{\smallskip}
4     &  {\footnotesize BMR}  & $10^{-2}$& $5\times10^{-3}$&  $\surd$       & $\surd$       & 2.91       \\
5     &  {\footnotesize BR}   & $10^{-2}$& $5\times10^{-3}$&  $\surd$       & $\times$      & No         \\
6     &  {\footnotesize BM}   & $10^{-2}$& $5\times10^{-3}$&  $\times$      & $\surd$       & 3.04       \\
  \hline\noalign{\smallskip}
7     &  {\footnotesize CMR}  & $10^{-3}$& $5\times10^{-3}$&  $\surd$       & $\surd$       & 1.79       \\
8     &  {\footnotesize CR}   & $10^{-3}$& $5\times10^{-3}$&  $\surd$       & $\times$      & No         \\
9     &  {\footnotesize CM}   & $10^{-3}$& $5\times10^{-3}$&  $\times$      & $\surd$       & 1.79       \\
  \noalign{\smallskip}\hline
\end{tabular}
\ec
\tablecomments{0.86\textwidth}{ The model names indicate different physics. A, B, and C indicates different $f_g$; M is the mass outflow feedback;
R is radiation pressure feedback. a: $f_g$ is the gas fraction. b: $\epsilon_w$ is the wind efficiency.
c: the time when the outer shell first arrives at $r=1$ kpc; and `No' means never.}
\end{table}

\section{Results}
\label{sect:results}
The gravity of the SMBH is only effective in the innermost regions. We set the mass of the SMBH to be $M_{BH}=10^8 \rm{M_\odot}$ and the velocity dispersion $\sigma=200\rm~km~s^{-1}$, which is consistent with the $M_{BH}-\sigma$ relation (see, e.g., \citealt{Ferrarese+etal+2000}). We assume the X-ray flux fraction $f_\ast=0.05$ and the UV flux fraction $f_d=1-f_\ast=0.95$. The gas fraction $f_g$ is taken as a free parameter. We calculated models with various $f_g$ and with each feedback term turned on/off. The models are summarized in Table \ref{tab:model_sum}. The gas fractions are given in Col. (3) and the wind efficiencies are given in Col. (4). The symbols ``$\surd$~{\rm /}$~\times$,'' shown in Cols. (5) and (6), indicate whether radiation pressure or mass outflow feedback is considered in the simulations. Col. (7) gives the time when the outer shell first arrives at 1 kpc.

\subsection{Triggering of star formation} \label{subsec:sf}
\begin{figure}
\centering
\includegraphics[width=7.25cm, angle=0]{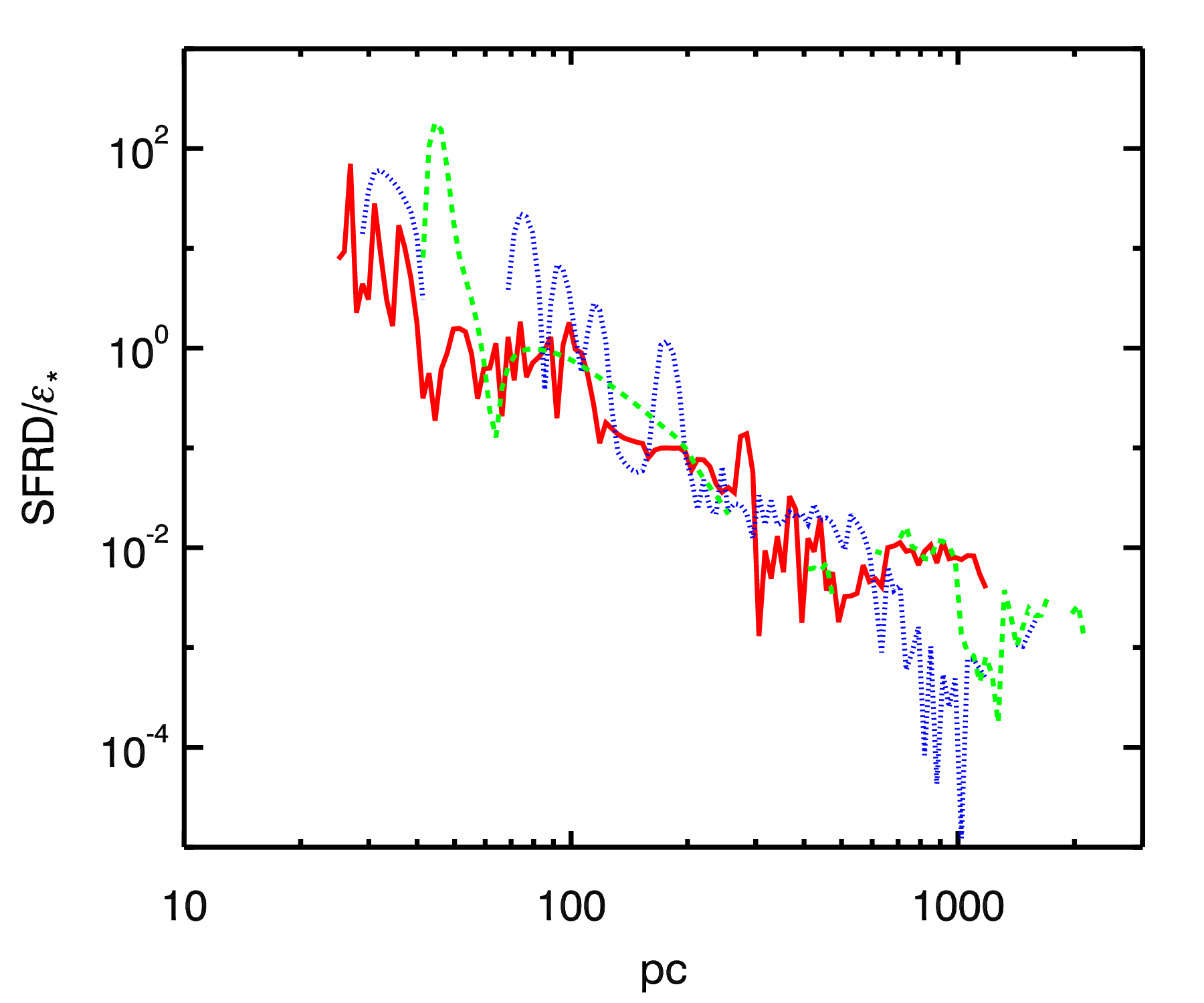}
\includegraphics[width=7.25cm, angle=0]{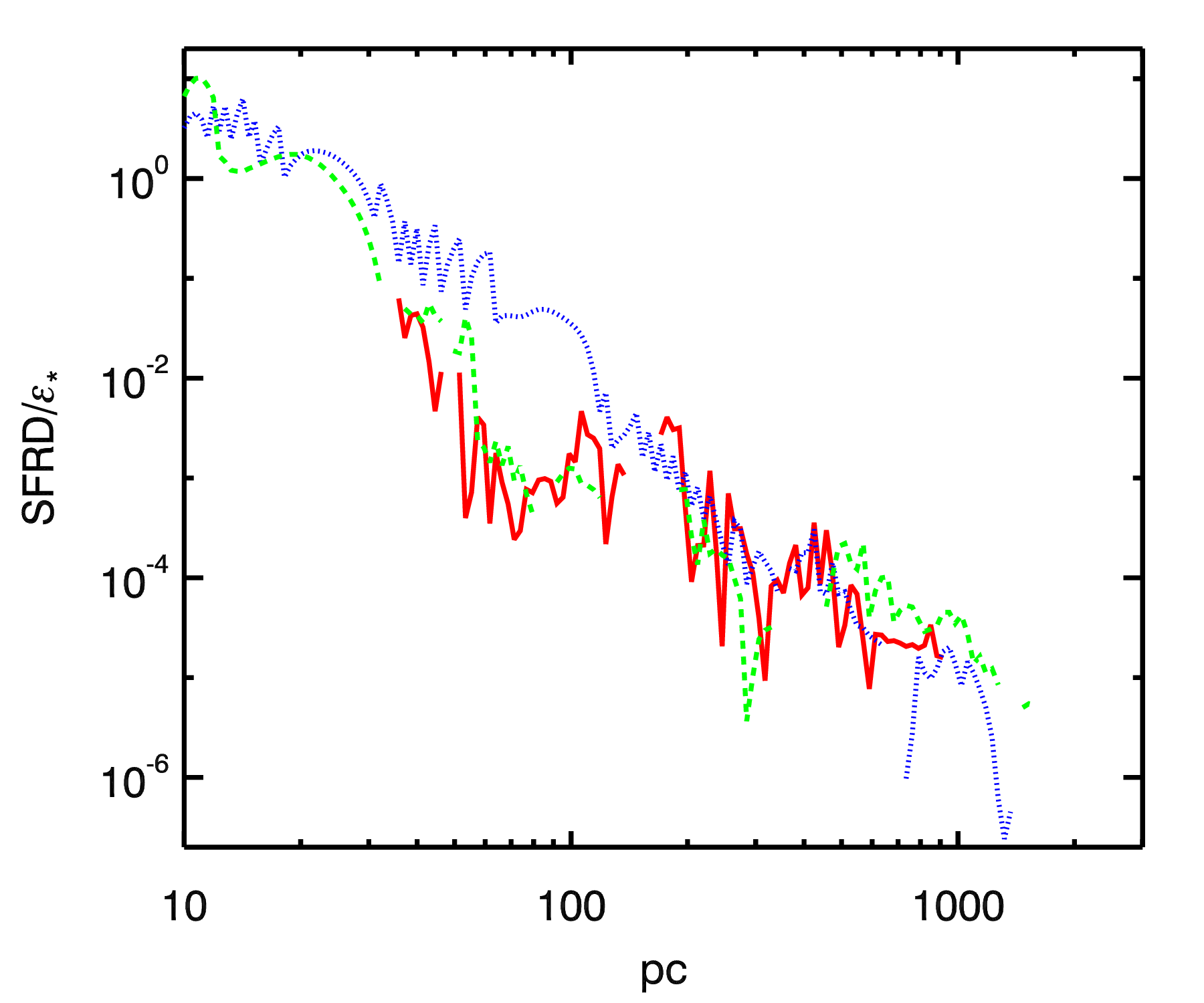}
\caption{{\small The radial profiles of angle-averaged SFRD ($\dot\rho_\ast$) for the AMR (left panel) and CMR (right panel) models. The solid, dotted and dashed lines correspond to the snapshots taken at t=2.235, 4.47 and 6.705 Myr, respectively. The vertical axis is $\dot\rho_\ast/\epsilon_*$ in units of ${\rm M_\odot yr^{-1} (100pc)^{-3}}$. The horizontal axis is in units of pc. We only include the dense gas with $T<10^5$ K in the calculation.}}
\label{fig:Fig2}
\end{figure}

From left to right columns, Figure~\ref{fig:Fig1} successively shows the logarithmic density, temperature and velocity magnitude contours of the ambient gas shocked by the mass outflow and radiation pressure. The top panels correspond to the gas-rich AMR model ($f_g=0.16$), and the bottom panels to the gas-poor CMR model ($f_g=10^{-3}$). The arrows in the right columns show the directions of the velocities. All of the panels in Figure~\ref{fig:Fig1} are snapshots taken at time $t=4.47$ Myr. Under the impact of the radiation pressure and ram pressure of the mass outflows, the inner interstellar medium is shocked into shells. Some of these shells fragment into clumps and filaments due to the Rayleigh-Taylor and thermal instabilities (cf. thin shell instability, which occurs when the radiative cooling is very strong, see \citealt{Vishniac+1983} and also \citealt{MacLow+etal+1989}). The clumps and filaments can be easily found in the left and middle panels; they have the highest density and lowest temperature. This phenomenon was also found by \cite{Nayakshin+etal+2012}. The cold, dense clumps and filaments, which survive the crash of the AGN feedback, are ideal places for star formation. We use a simple prescription to evaluate the star formation rate (SFR) as adopted by \cite{Ishibashi+etal+2012}, and the formula is
\begin{equation}\label{eq:sfr}
   \dot M_\ast \sim \epsilon_\ast\frac{M_g}{t_{flow}} \sim \epsilon_\ast\frac{2f_g\sigma^2}{G}v({\bf r}),
\end{equation}
where $\epsilon_\ast$ is the star formation efficiency, $M_g$ is the gas mass and $t_{flow}=r/v({\bf r})$ is the local flow time. The observed star formation efficiency $\epsilon_\ast$ is $\sim0.01-0.1$. For the AMR model, we obtain the SFR at large radii $\dot M_{\ast,\infty}\sim 7.2\times 10^{0-1}~{\rm M_\odot yr^{-1}}$ if $v_\infty=60~{\rm km~s^{-1}}$ (see top-left panel of Figure \ref{fig:Fig1}). By analogy, we estimate the star formation rate density (SFRD) by using the formula $\dot \rho_\ast\sim \epsilon_\ast \rho/t_{flow}$. The radial profiles of the angle-averaged $\dot \rho_\ast/\epsilon_\ast$ are plotted in Figure \ref{fig:Fig2} for the AMR (left panel) and CMR (right panel) models with the solid, dotted and green lines corresponding to the snapshots taken at t=2.235, 4.47 and 6.705 Myr, respectively. The vertical axis is in units of ${\rm M_\odot yr^{-1} (100pc)^{-3}}$. From the left panel, we find that the SFRD around 1 kpc at 4.47 Myr is $\sim 10^{-4}\epsilon_\ast~{\rm M_\odot yr^{-1} (100pc)^{-3}} = 10^{6} (\epsilon_\ast/0.01)~{\rm M_\odot yr^{-1} (Mpc)^{-3}}$. The largest SFRD in the cosmological simulations, including the AGN feedback presented by \cite{Fanidakis+etal+2012}, is $\sim 0.1{\rm~M_\odot yr^{-1} (Mpc)^{-3}}$ (see their Figure 1), which is several orders of magnitude smaller than our estimation. The inconsistency is caused by several reasons. One of the most important reasons is that we ignore the fact that most of the protostars and formed stars will be disrupted when the clumps, filaments
and shells move away from the center of the galaxy (\citealt{Nayakshin+etal+2012}). Another reason is that the SFRD we obtain from our simulation domain only covers the galactic center, but the SFR is usually inversely proportional to the radius, so the SFRD will inevitably be overestimated if we extrapolate our results directly to a large scale. These details will be studied more accurately in the future.

Clumps and filaments also appear in the gas-poor CMR model, which is a new result that was not found by \cite{Nayakshin+etal+2012}.  This is because we have line radiation pressure and more complete radiative heating/cooling, so the thin shell instability occurs although it is much weaker for the gas-poor case. The consequent SFRD is much smaller compared with the gas-rich case (see the right panel of Figure \ref{fig:Fig2}). For example, SFRD around 1 kpc at 4.47 Myr is $\sim 10^{-5} \epsilon_\ast~{\rm M_\odot yr^{-1} (100pc)^{-3}}$, which is one order of magnitude smaller than the gas rich case. It should be pointed out that, since the feedback is weaker when the ambient gas density is smaller under our model construction, the outward velocity of the gas shell is much smaller compared with simulation F0.03 in \cite{Nayakshin+etal+2012}.

\begin{figure}
\centering
\includegraphics[width=13.5cm, angle=0]{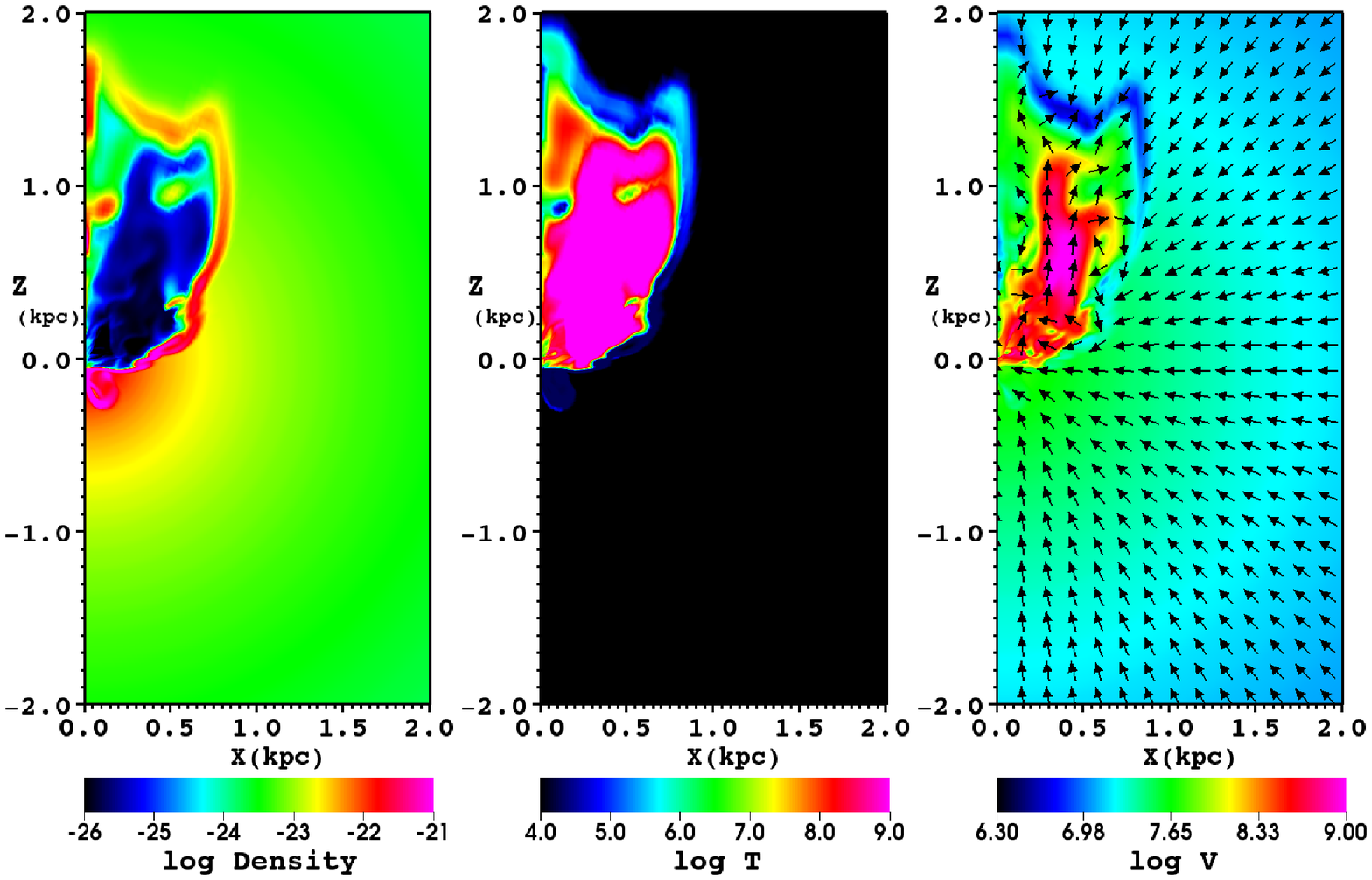}
\includegraphics[width=13.5cm, angle=0]{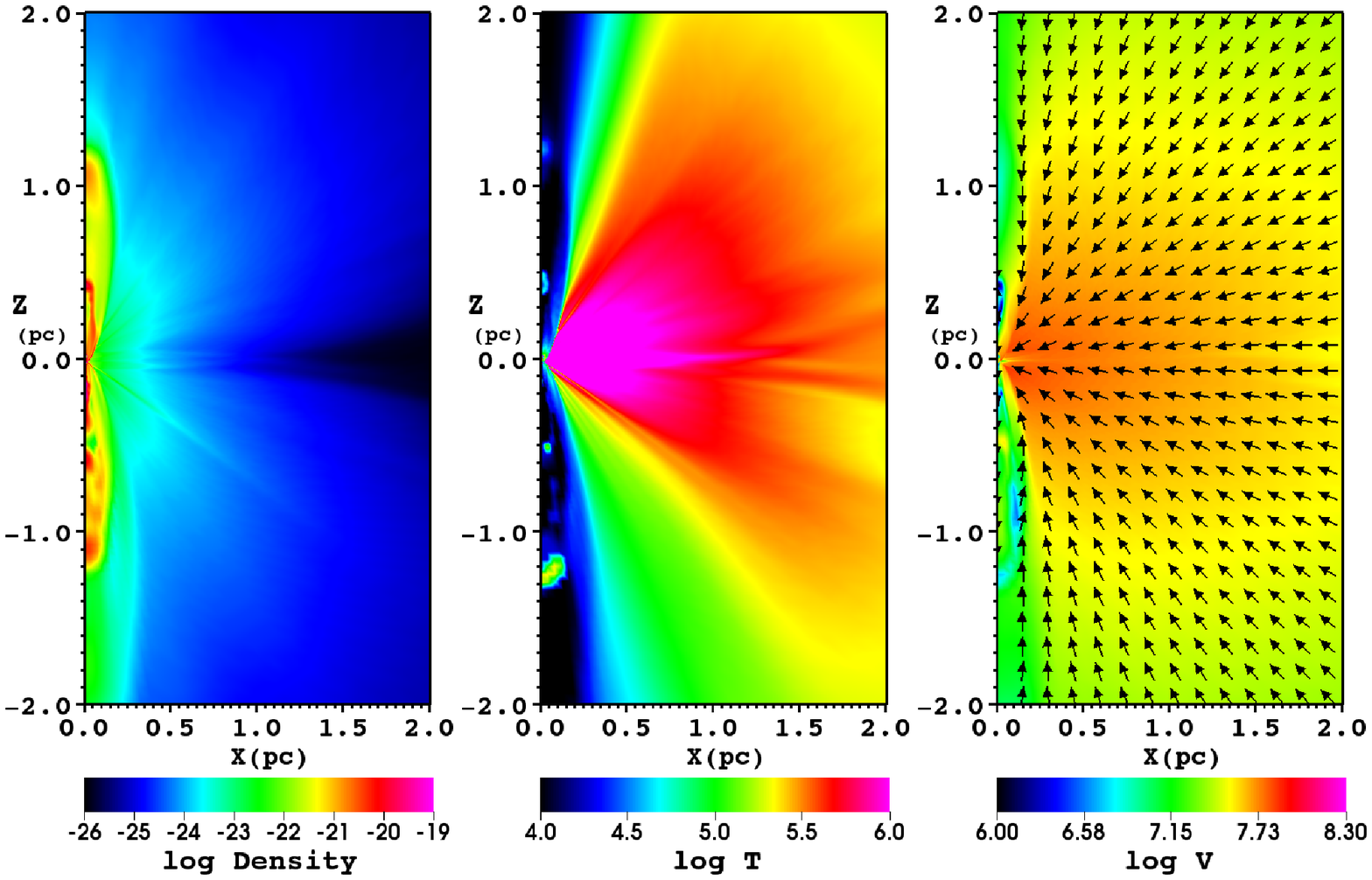}
\caption{{\small Top panels: the AM model without radiation pressure feedback; bottom panels: the AR model with only radiative feedback. The physical quantities for the panels are the same as in Figure~\ref{fig:Fig1}. The top panels are snapshots taken at 4.47 Myr, and the bottom panels present
the snapshots of the AR model at 27.71 Myr.}}
\label{fig:Fig3}
\end{figure}

\subsection{Gas-rich galaxies: mutually reinforced feedback} \label{subsec:gasrich}
In this and the next subsection, we explore the feedback effects by turning on/off one of the two forms of feedback. Radiative heating/cooling is always turned on. Firstly, we study the gas-rich case.

Figure~\ref{fig:Fig3} shows the same physical quantities as Figure~\ref{fig:Fig1} but for the AM (top panels, without radiation pressure feedback) and AR (bottom panels, with only radiative feedback) models. Radiative heating/cooling is always turned on. The top panels of the AM model are snapshots at time 4.47 Myr, the bottom panels present snapshots of the AR model at time 27.71 Myr. Both the AM and AR models have much weaker feedback effect with respect to the smaller effective feedback radii, by having fewer clumps and filaments compared with the AMR model. For the AM model without radiation pressure feedback, the shocked shells are not symmetrical about the middle plane with $z=0$. This is because at the very beginning, the shocked shell above the $z=0$ plane fragments first, then the outflow and radiation traverse the gaps of the fragmented shell and interact with the outer ISM, and then the upper bubble grows much more quickly. On the other hand, some of the gas that is blown away in the upper panel would fall back down. It would be harder to blow away the gas below the equatorial plane. Therefore, once the asymmetry appears, the growth in asymmetry will be a runaway process. We have not shown the snapshots of the AR model at time 4.47 Myr because the feedback effects are only effective in the very inner regions. The luminosity we use is designed to produce sufficient radiation pressure to balance the gravity. But why does the radiation pressure not seems large enough? The reason is that the SMBH's gravity is not included in the luminosity calculation. After long time evolution, most of the gas flows into the inner boundary along the mid-plane. We can see from the bottom panels of Figure \ref{fig:Fig3} that the radiation pressure feedback only has an influence along the polar axis and in small regions even after long time evolution.

We can see that the regions where effective feedback occurs are almost inside the outer shell from the top panels of Figure \ref{fig:Fig3}. As shown in the bottom panels in the same figure, the gas temperature can be above $10^6$ K at large radii, which is a result of radiative heating, and means that the regions where effective feedback occurs are well beyond the outer shell. The discrepancy between these two models is due to the fact that the gas density of the outermost shell is so high in the AM model that the shell is optically thick, but the outermost shell in the AR model is still optically thin so the photons from the central engine can penetrate the shell to heat the gas from a distance. \cite{Ciotti+etal+2010c} claimed a similar conclusion, the radiative heating is more effective than the mass outflow feedback at relatively much larger radii. In comparison with the AMR model which includes both feedbacks, the feedback effects in the AM and AR models are much weaker. In other words, the two types of feedbacks reinforce each other. This is because the long-distance radiative feedback can accelerate gas at large radii, which reduces the resistance so that the mass outflows more easily propagate forward to impact the ISM at large radii. On the other hand, when the adjacent gas is compressed by mass outflows, the line force would increase if the temperature is lower than $10^5$ K. In general, the physical mechanisms included here are very complicated. We will leave detailed analysis for future work.

\begin{figure}
\centering
\includegraphics[width=13.5cm, angle=0]{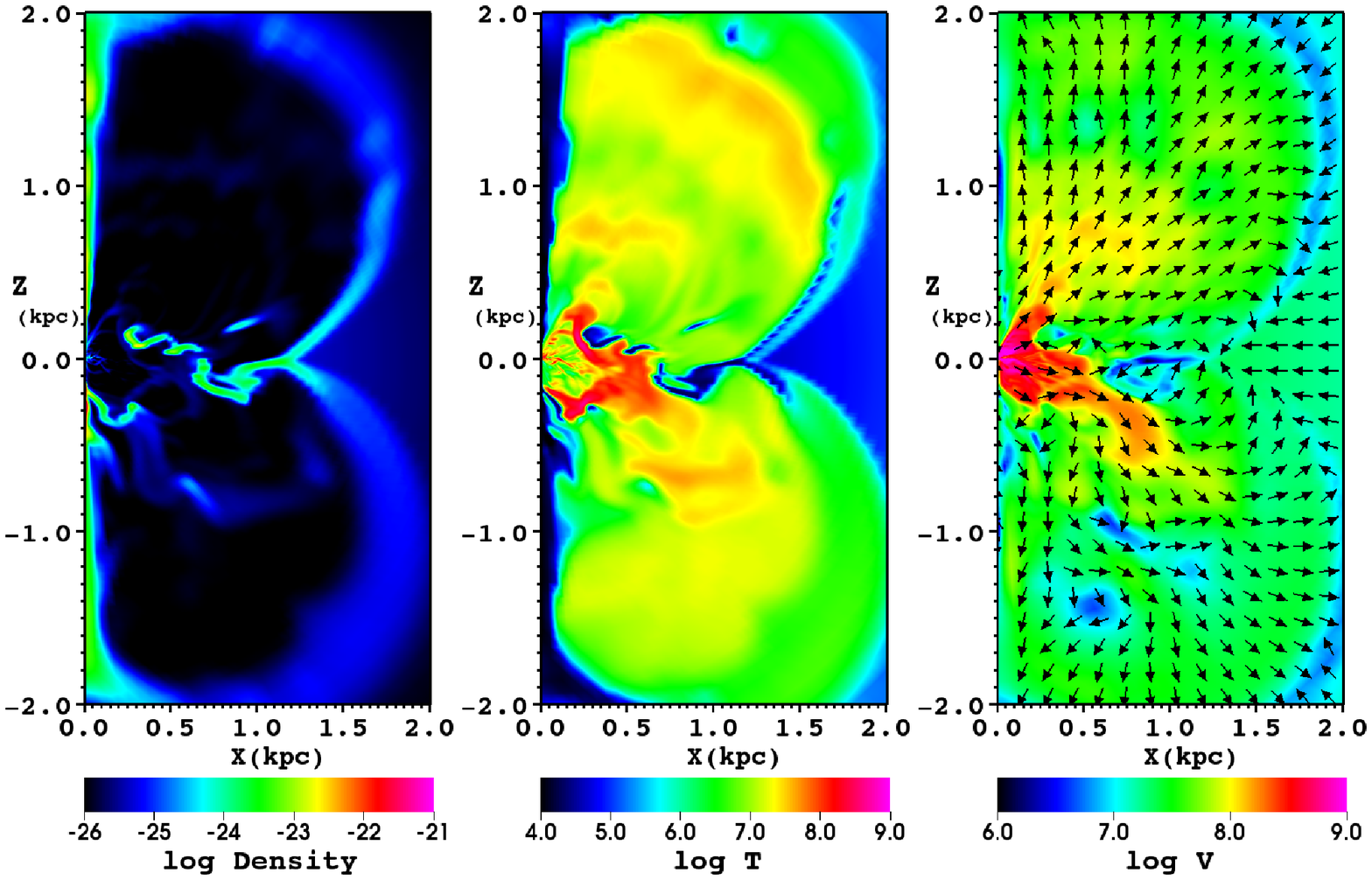}
\includegraphics[width=13.5cm, angle=0]{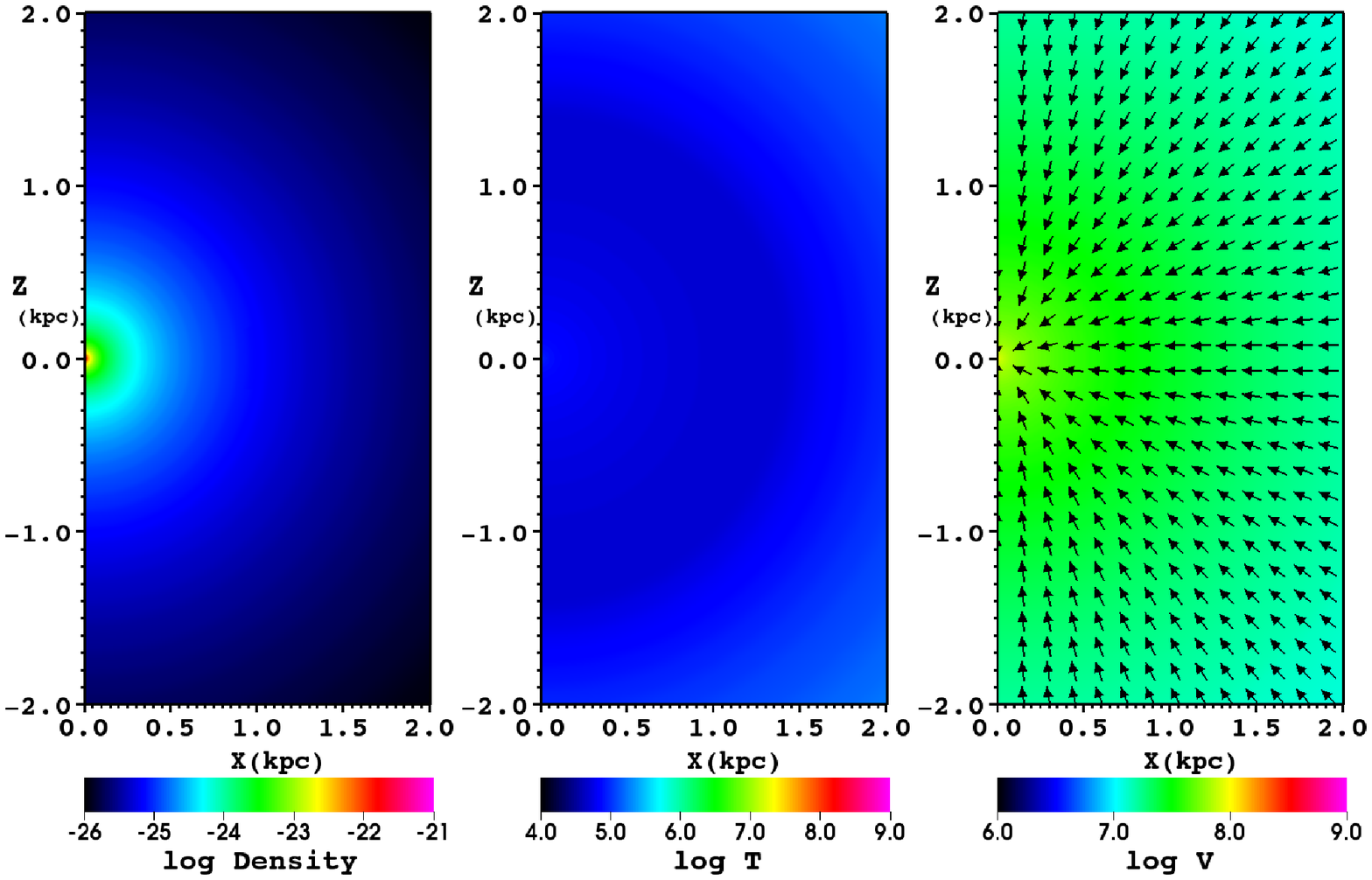}
\caption{{\small Top panels: the CM model without radiation pressure feedback; bottom panels: the CR model with only radiative feedback. The physical quantities for the panels are the same as in Figure~\ref{fig:Fig1}. All the panels are snapshots taken at 4.47 Myr.}}
\label{fig:Fig4}
\end{figure}

\subsection{Gas-poor galaxies: outflow feedback dominated} \label{subsec:gaspoor}
We turn to the gas-poor case in this subsection. The results of the CM and CR models are shown in Figure \ref{fig:Fig4}. The figure pattern is the same as in Figure \ref{fig:Fig3}, except that all the snapshots are taken at time 4.47 Myr. For model C series, the luminosity $L_{acc}\approx 2.2\%L_{Edd}$, which is too small to produce effective radiative feedback in our current models. There are no radiation-driven outflows in the CR model as seen in the bottom panels of Figure \ref{fig:Fig4}. However, the mass outflow feedback is very efficient in sweeping off the ISM. Comparing the top-left panel of Figure \ref{fig:Fig4} with the bottom-left panel of Figure \ref{fig:Fig1}, we conclude that mass outflow feedback is dominant in the gas-poor case. The more complicated structures of the CMR model indicate that radiation pressure also plays a role in the processes of feedback-ISM interactions.

Until now, we have found that both radiation pressure and mass outflow are important to the evolution of the ISM, and that the associated feedback processes are very complicated.

\section{Conclusions and Discussion}
\label{sect:concl}
We perform 2D HD numerical simulations considering both radiative and mass outflow feedbacks to study positive AGN feedback. We take a more intact treatment of radiative heating/cooling including Compton heating/cooling, photoionization heating/recombination cooling, bremsstrahlung and line cooling. Besides Thomson scattering, we also consider line-absorption. This is by far one of the most sophisticated simulations on the topic of positive AGN feedback. We primarily evaluate the SFR and SFRD in the gas shells, clumps and filaments which are believed to be ideal places for star formation. The clumps and filaments are generated by the fragmentation of gas shells. We find that the SFR is greatly enhanced rather than disrupted by AGN feedback, as also found by \cite{Nayakshin+etal+2012}. And we also find that star formation is triggered even in gas-poor case. Furthermore, we find that, although radiation pressure feedback has a limited effect, when mass outflow feedback is also included, they reinforce each other. We conclude that AGN feedback can be positive with respect to increasing SFR, and that both radiation and mass outflow play important roles in this SMBH-galaxy co-evolution scenario.

\begin{figure}
\centering
\includegraphics[width=14cm, angle=0]{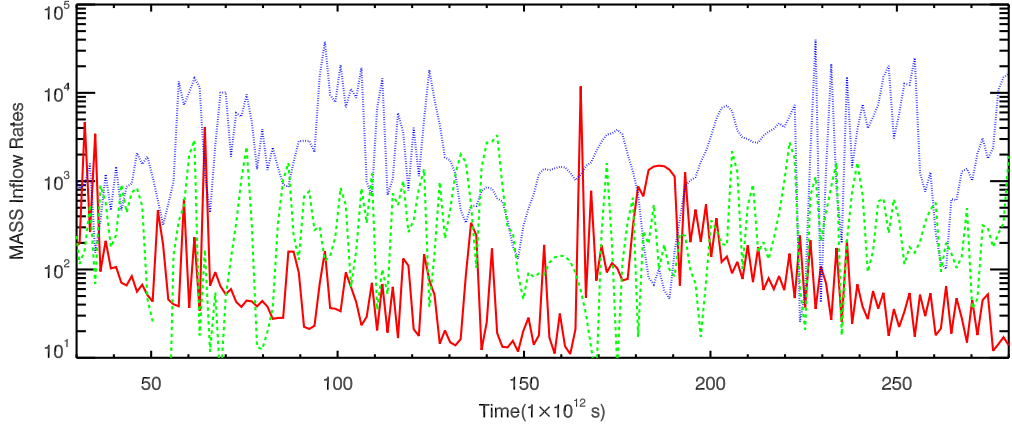}
\caption{{\small The time evolution of mass inflow rates for the AMR (red, solid line), BMR (blue, dotted line) and CMR (green, dashed line) models.
The vertical axis is in units of $10^{25}~\rm{g~s^{-1}}=0.625~\rm{M_\odot~yr^{-1}}$, and the horizontal axis is in units of $10^{12}~\rm{s}=3.17\times10^4~\rm{yr}$.}}
\label{fig:Fig5}
\end{figure}

We explore the dependence on ISM density by varying the gas fraction $f_g$, as summarized in Table \ref{tab:model_sum}. We do not derive the results of the BMR model because it is an intermediate model between the AMR and CMR models. We set the luminosity proportional to $f_g$ so that the gas-rich models have a higher luminosity and vice versa, and we find that for the high luminosity models, both radiation and mass outflow feedback are important as they reinforce each other. For models with low luminosity, it is hard for the radiation pressure to drive strong outflows, although it is still effective. That is to say, the models with low luminosity are dominated by mass outflow feedback. These conclusions are reasonable if we take the standard thin disk (\citealt{Shakura+etal+1973}) or slim disk (\citealt{Abramowicz+etal+1988}) to explain the high luminosity models, and take the radiation inefficient accretion flow (RIAF; \citealt{Narayan+etal+1994}) to explain the low luminosity models. Mass outflows have been found in both the luminous standard thin disk (e.g., \citealt{Murray+etal+1995}; \citealt{Proga+etal+2000}), slim disc \citep[e.g.,][]{Ohsuga+etal+2005, Ohsuga+etal+2011} and faint RIAF \citep[e.g.,][]{Yuan+etal+2012b, Yuan+etal+2012a} in the past 15 yrs or so. We calculate the mass inflow rates for the AMR, BMR and CMR models, and their time evolutions are drawn in Figure \ref{fig:Fig5}. The mass inflow rates are very large and highly time varying. The reason for the variation in time is that some of the gas that is blown-away falls back again, especially around the mid-plane.

\cite{Nayakshin+etal+2012} did not evaluate AGN feedback triggered SFR. They showed that the pressure in the shocked ambient gas is much larger than the maximum gas pressure in the pre-quasar host. Physically, by compressing the cold gas in the hosts, the strong pressure is able to trigger star formation or enhance it. \cite{Ishibashi+etal+2012} give SFRs on the order of $\sim 10-100~{\rm M_\odot~yr^{-1}}$ for $\epsilon_\ast\sim 0.01-0.1$ and $f_g=0.16$. This value is comparable to our results for the AMR model. The ignored dust effects have several aspects: one is that the dust opacity can enhance the radiation pressure; another effect is that the dust absorb the UV photons and then re-emit in the IR band; finally, the dust is tightly related with the star formation processes. Until now, no studies have self-consistently calculated the SFR, although most claim that the conditions for star formation are satisfied. We emphasize that implementing the star formation processes into the simulations is necessary.

There are some other caveats we should keep in mind. Our treatment of accretion luminosity is not self-consistent. To correctly capture accretion, feedback and star formation in a single simulation is a very formidable project. However, we can treat star formation in a semi-analytic way (\citealt{Ciotti+etal+2007}), and parameterize the relation between the accretion rate onto the black hole and the mass inflow rate at the inner boundary. To fully understand radiative feedback, especially when dust is considered, radiation transfer simulation is needed. These improvements will be studied in future works.

\normalem
\begin{acknowledgements}
We thank Prof. Feng Yuan for his useful suggestions, and Dr. Tomohisa Kawashima for his thoughtful comments/suggestions.  We are also very grateful to the anonymous referee for his/her informative and instructive comments. This work was supported in part by the National Natural Science Foundation of
China (grants 11203057, 11103061, 11133005), and the Shanghai Postdoctoral Scientific Program (grant 11R21417700). The simulations were carried out at the Shanghai Supercomputer Center.

\end{acknowledgements}

\bibliographystyle{raa}
\bibliography{ms1419bibtex}

\end{document}